\documentclass[letterpaper,english,prl,final,twocolumn]{revtex4}
\usepackage[latin9]{inputenc}
\usepackage{units}
\usepackage{amsmath}
\usepackage{graphicx}
\usepackage{amssymb}

\makeatletter
\@ifundefined{textcolor}{}
{%
 \definecolor{BLACK}{gray}{0}
 \definecolor{WHITE}{gray}{1}
 \definecolor{RED}{rgb}{1,0,0}
 \definecolor{GREEN}{rgb}{0,1,0}
 \definecolor{BLUE}{rgb}{0,0,1}
 \definecolor{CYAN}{cmyk}{1,0,0,0}
 \definecolor{MAGENTA}{cmyk}{0,1,0,0}
 \definecolor{YELLOW}{cmyk}{0,0,1,0}
 }

\makeatother

\usepackage{babel}

\begin{document}

\title{{\Large Nuclear Scission and Quantum Localization}}

\author{W. Younes and D. Gogny}

\affiliation{Lawrence Livermore National Laboratory, Livermore, CA 94551}

\date{\today}
\begin{abstract}
We examine nuclear scission within a fully quantum-mechanical microscopic
framework, focusing on the non-local aspects of the theory. Using
$^{240}\textrm{Pu}$ hot fission as an example, we discuss the identification
of the fragments and the calculation of their kinetic, excitation,
and interaction energies, through the localization of the orbital
wave functions. We show that the {}``disentanglement'' of the fragment
wave functions is essential to the quantum-mechanical definition of
scission and the calculation of physical observables. Finally, we
discuss the fragments' pre-scission excitation mechanisms and give
a non-adiabatic description of their evolution beyond scission.

\end{abstract}
\maketitle
Nuclear scission, the process wherein a nucleus breaks into two or
more fragments, poses a fundamental challenge to quantum many-body
theory: scission implies a separation of the nucleus into independent
fragments, while the Pauli exclusion principle introduces a persistent
correlation between the fragments, no matter how far apart they are.
The objective of this paper is to resolve this paradox by disentangling
the fragments in a fully quantum-mechanical description that is consistent
with experimental data. In addition to shedding light on fundamental
aspects of many-body physics, a microscopic theory of scission is
needed to make reliable predictions of fission-fragment properties,
such as their excitation and kinetic energies, and their shapes. In
particular, we revisit in a microscopic approach the question of the
energy partition between light and heavy fragments which was addressed
in a recent letter \cite{schmidt10} within a statistical-mechanic
treatment. While many technical challenges remain in the 70-year quest
to develop a predictive theory of fission, understanding scission,
remains a formidable conceptual obstacle to such a theory.

Previous descriptions of scission have always been formulated within
the context of a nuclear density, with an identifiable neck joining
two pre-fragments. The neck ruptures at some point along its length,
and all the matter to one side or the other of the rupture is relegated
to the corresponding fragment. Despite its usefulness, this is ultimately
a classical view of scission. In 1959 \cite{whetstone59}, this picture
was used to qualitatively account for the different observed mass
divisions in fission and the well-known {}``sawtooth'' shape of
the average neutron-multiplicity distribution. Later on, a more quantitative
description of the nuclear shape was introduced \cite{nix69}, and
scission was equated with a vanishing neck size. This criterion was
later improved \cite{davies77} by requiring that scission occurs
when the Coulomb repulsion exceeds the attractive nuclear force between
the fragments. Nörenberg \cite{norenberg72} took a step toward a
more microscopic description using a molecular model of fission calculated
in a two-center Hartree-Fock+BCS approach. Bonneau \textit{et al}.
\cite{bonneau07} used separate microscopic calculations of each fragment
and a phenomenological nuclear interaction between them to define
a scission criterion based on the ratio of their mutual nuclear and
Coulomb energies. In recent calculations \cite{goutte05,dubray08,younes09}
the entire fissioning nucleus was treated within a single microscopic
framework and the properties of the nucleus at scission were calculated.
In those calculations however, the identification of scission and
calculations of fragment properties still relied on the nuclear density.
In contrast to previous approaches, we present here a fully quantum-mechanical
description of scission that accounts for the nonlocality of the many-body
wave function of the nucleus. The need for, and difficulty of disentangling
the fragment wave functions was alluded to in \cite{bonneau07}. Our
solution is in the spirit of the Localized Molecular Orbital (LMO)
technique used in molecular physics \cite{lennardjones49}: we localize
individual orbitals on the fragments while the nucleons themselves,
described by a Bogoliubov vacuum built from these states, remain delocalized.
This powerful technique has never been used to describe nuclear scission
before.

The work described in this paper is based on constrained Hartree-Fock-Bogoliubov
(HFB) calculations of $^{240}\textrm{Pu}$ with a finite-range (D1S)
interaction. Details of the calculation are given in \cite{younes09}.
We have chosen to focus on the hot-scission point with constrained
quadrupole moment $Q_{20}$ = 370 b \cite{berger89}, and used the
constraint on neck size, $Q_{N}$, to approach scission. This constraint
lets us vary the density of matter in the neck. HFB calculations produce
self-consistent solutions that minimize the total energy of the nucleus.
In order to describe the nucleus near scission, we introduce here
the additional requirement that the interaction energy between pre-fragments
must be minimized. This criterion is consistent with the physical
picture of the pre-fragments evolving into independent fragments that
move increasingly further apart. We have shown in previous work \cite{younes09,younes09a}
that the pre-fragments in the HFB solutions near scission generally
exhibit {}``tails'', portions of individual orbital wave functions
that extend into the complementary fragment. Our calculations show
that the size of these tails is closely related to the strength of
the interaction between the fragments. Therefore, minimization of
the interaction energy between fragments is essentially equivalent
to localization of the orbitals on the fragments.

\begin{figure}[htbp]
\centering{}\includegraphics[scale=0.31]{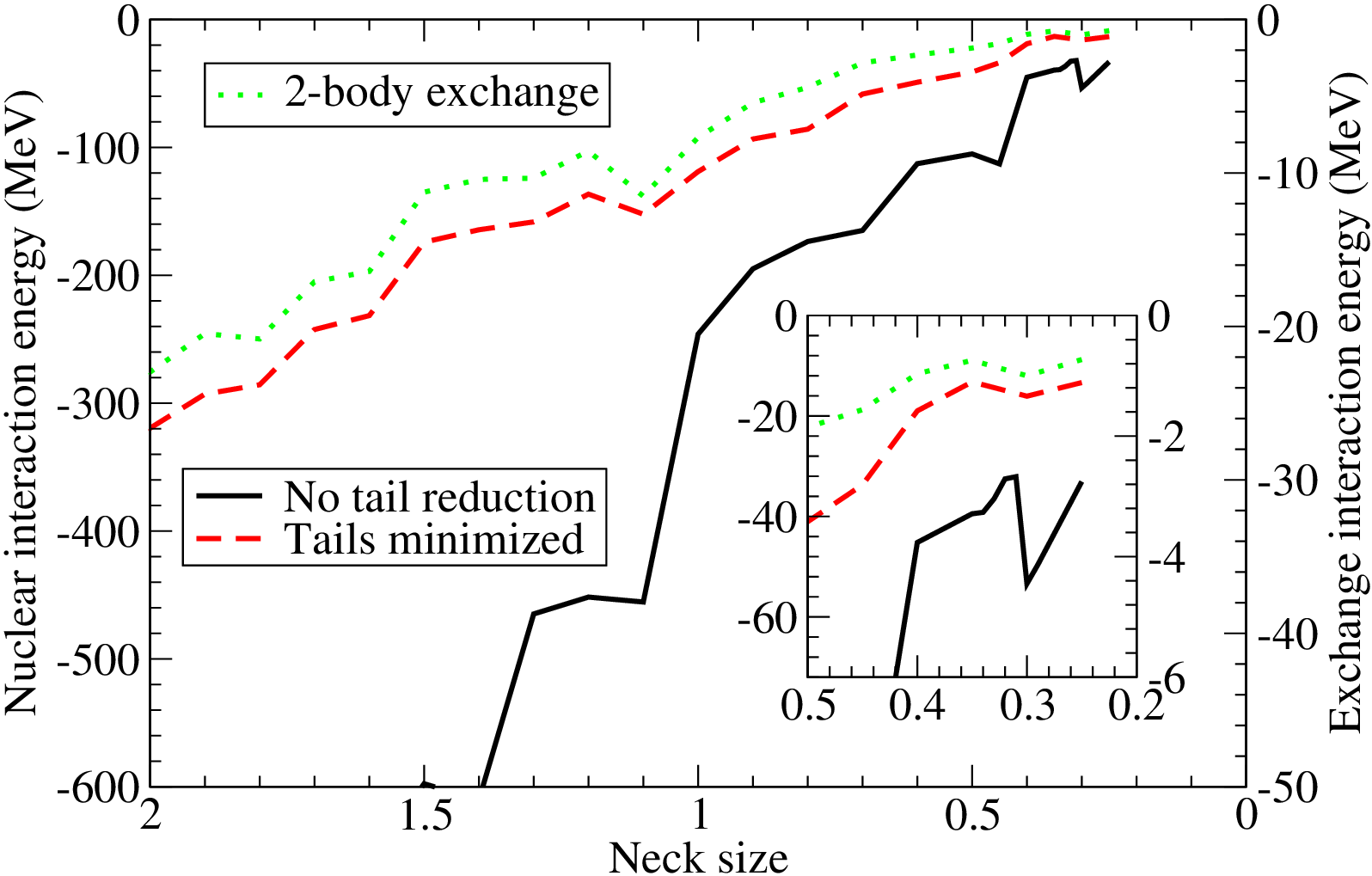} \caption{\label{fig:eint}(Color online). Interaction energies plotted as a
function of neck size ($Q_{N}$). The solid black and red dashed curves
are the nuclear interaction energies before and after localization
respectively (energy scale on left y axis), and the dotted green curve
is the exchange part of the 2-body component of the interaction energy
(energy scale on right y axis). The inset shows a closeup view for
$0.2\leq Q_{N}\leq0.5$.}

\end{figure}

Hartree-Fock methods in molecular (or nuclear) physics generally produce
single-particle atomic orbitals that are not spatially localized within
a molecule (nucleus). However, it was observed early on \cite{lennardjones49}
that any unitary transformation applied to the single-particle components
of a Slater determinant does not affect the global properties of the
corresponding system. Since then, unitary transformations have been
routinely used to localize electron orbitals and thereby define such
chemically meaningful concepts as core and bond orbitals. In nuclear
fission, we will use the same concept to localize nuclear quasi-particle
states (qp) on the nascent fragments, taking advantage of the fact
that the Bogoliubov vacuum is only defined up to a unitary transformation
of the qp destruction operators. More precisely, for each qp \textit{$i$},
we define a localization indicator $\ell_{i}$, as the absolute difference
between the qp density to the left and right of the neck position.
A value $\ell_{i}=0$, for example, corresponds to a completely delocalized
qp. For a given pair of of qp, $\left(i,j\right)$, we can then look
for a mixing angle $\theta$ that maximizes the pair localization
parameter $\sqrt{\ell_{i}^{2}+\ell_{j}^{2}}$. Thus, through a systematic
search algorithm, a set of qp pairs is found that minimizes the summed
tail size of the two fragments. In selecting these pairs, we have
required that the level energies of the qp pairs are no more than
2 MeV apart, and have taken care not to mix {}``mirror'' states
(nearly degenerate in energy, but very different quasiparticle occupation
numbers). The interest of this process is that it unambiguously identifies
pre-fragments built from qps that are spatially localized. The interaction
energy between fragments can now be rigorously calculated as those
contributions to the HFB mean and pairing fields that couple qps in
complementary fragments. The nuclear component of that interaction
energy (i.e., excluding the direct Coulomb repulsion between fragments)
is plotted in Fig. \ref{fig:eint} before and after tail reduction.
In both cases individual qps are assigned to one fragment or the other
based on their spatial localization relative to the neck position.
The effect of the tail reduction can be rather substantial even when
the neck between the fragments is small, e.g. by $\sim20$ MeV even
when $Q_{N}$ \textless{} 0.5.

\begin{center}
\begin{figure}[htbp]

\centering{}\includegraphics[scale=0.34]{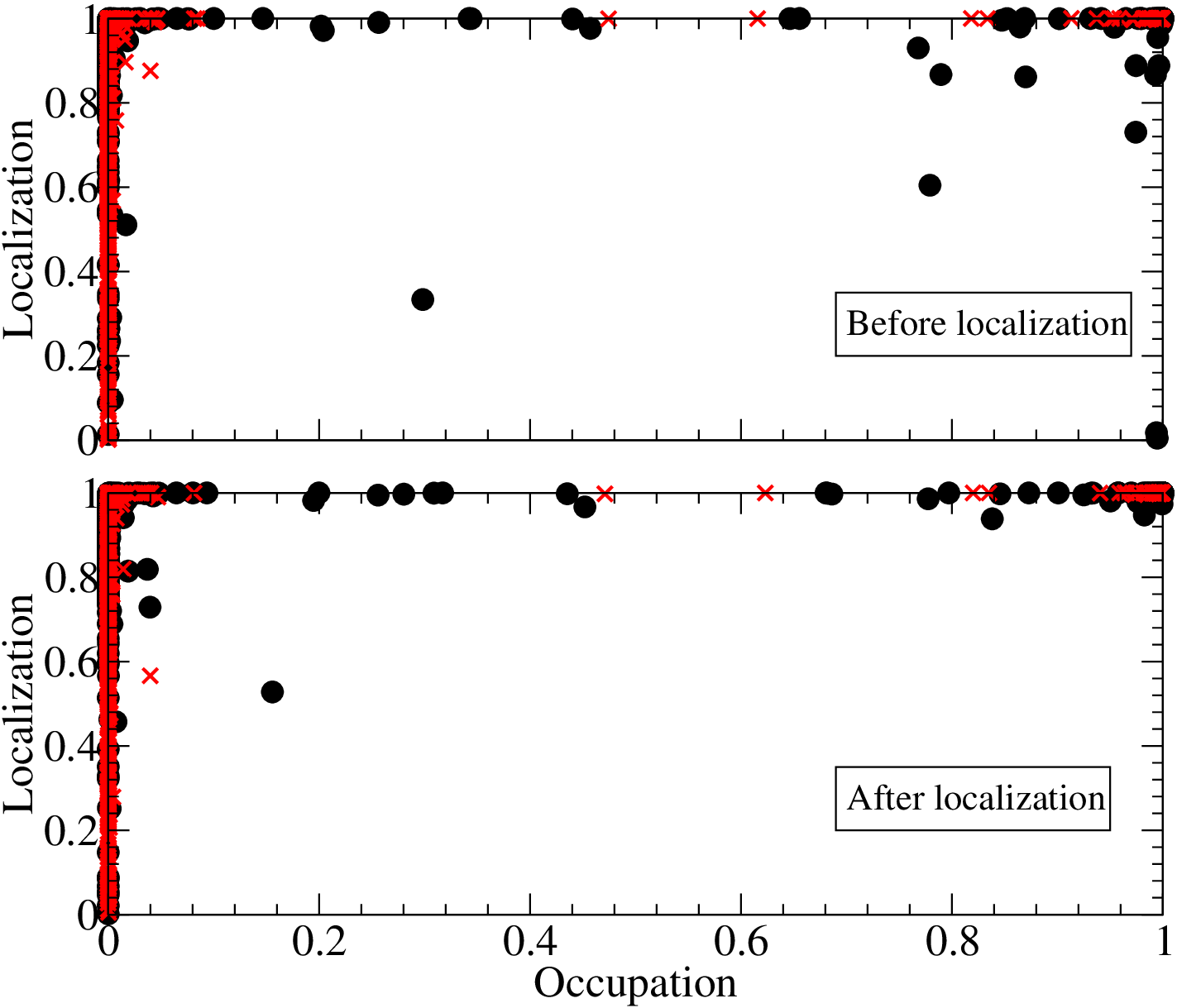} \caption{\label{fig:loc-ind}(Color online) individual quasiparticle states
before (top panel) and after (bottom panel) localization at scission.
Proton states are shown as red crosses, and neutron states as black
disks. The x axis gives the occupation ($v_{i}^{2}$) of the state,
while the y axis gives its normalized localization ($\nicefrac{\ell_{i}}{2v_{i}^{2}}$)
.}

\end{figure}

\par\end{center}

\begin{figure}[htbp]
\centering{}\includegraphics[bb=5bp 50bp 706bp 520bp,scale=0.34]{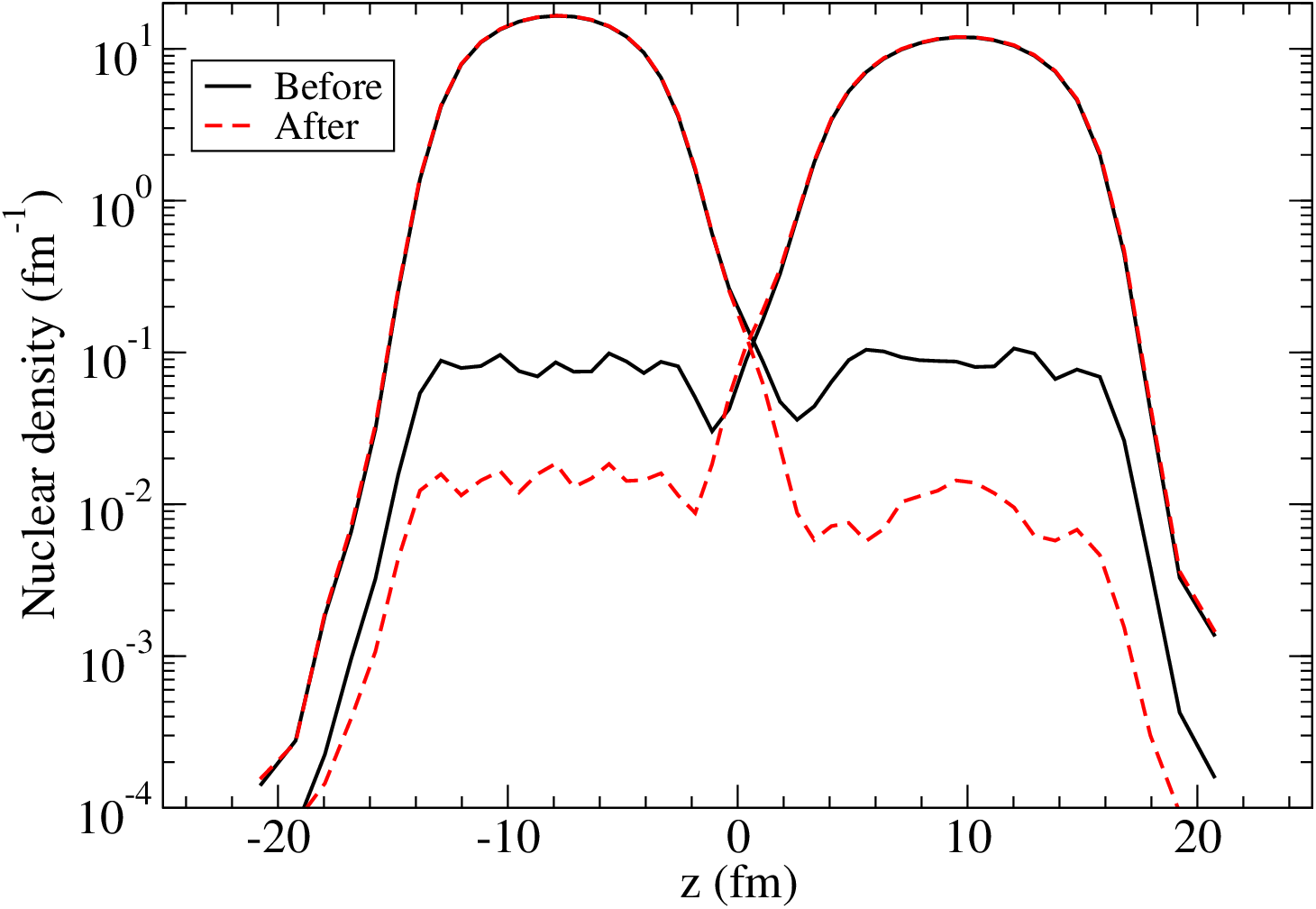}
\caption{\label{fig:loc-tot}(Color online) effect of localization on the densities
of the fragments at scission. The solid black line gives the density
along the symmetry (z) axis before localization, and the dashed red
line is the density after localization.}

\end{figure}

We show in Fig. \ref{fig:loc-ind} more details concerning the localization
of the qps with occupation $v^{2}$ according to whether they are
preferentially holes ($v^{2}$\texttt{\textgreater{}}1/2) or particles
($v^{2}$\texttt{\textless{}}1/2). We observe that the effect of the
localization is most visible for the hole states with $v^{2}$ \texttt{\textgreater{}}
0.7. Note in particular the pair of deeply-bound states in the top
panel of Fig. \ref{fig:loc-ind} with $v^{2}\approx$ 1 and $\ell\approx$
0 (i.e., fully delocalized), both with K quantum number 1/2 and only
7 keV apart in energy. These two states become fully localized in
the bottom panel. Notice also that a great number of localized qps
of particle type can combine with localized qps of hole type to provide
a rich spectrum of two-qp states localized on each of the two pre-fragments.
These simple excitations or combinations of them describe excited
fragments. Not all states are fully localized by the algorithm above,
in particular a 2-MeV, K = 1/2 neutron state remains in the bottom
panel with $v^{2}\approx$ 0.16 and $\ell\approx$ 0.53, but the overall
effect on the fragment densities shown in Fig. \ref{fig:loc-tot}
is significant. The effect of the localization on the interaction
energy is even more striking, as shown in Fig. \ref{fig:eint}. We
point out that this analysis includes $\approx1100$ proton and neutron
qp states.

If we faithfully apply the variational principle to the fissioning
nucleus in order to minimize the total energy (by the HFB method),
as well as the interaction energy (by tail reduction), the result
will be two infinitely separated fragments in their respective ground
states. Experimental observables--neutron emission and kinetic energies--clearly
indicate that we must depart from this adiabatic picture. In fact,
the point where the evolution of the fissioning nucleus ceases to
be adiabatic could be taken as a definition of scission. For practical
applications, we give the following 3 criteria that define the scission
point: 1) the repulsive Coulomb force between fragments greatly exceeds
their mutual nuclear attraction, 2) the exchange contribution to the
interaction between fragments is small, which means that we can neglect
the antisymmetry between their constituents and describe the system
as two separate Bogoliubov vacua, and 3) in each fragment, we can
excite a set of two-qp states that remain localized on the fragments,
so that the fragments can be considered as separate entities with
their own excitations and in interaction through a repulsive force
acting only on their respective centers of mass. As the neck is reduced
in our $^{240}\textrm{Pu}$ calculation, the point at $Q_{N}$ = 0.35
is the first for which all three criteria above are simultaneously
verified, and the results in Figs. \ref{fig:loc-ind} and \ref{fig:loc-tot}
were calculated for this scission point. Scission may occur at other
nearby points, but this one is representative. At $Q_{N}$ = 0.35,
the Coulomb force is $\approx$ 30 times larger than the nuclear one,
the two-body exchange contribution is only -0.7 MeV (Fig. \ref{fig:eint}),
and a set of 2-qp states can be constructed from localized states
in Fig. \ref{fig:loc-ind} that remain localized within a fragment
(i.e., their creation does not significantly affect the excitation
of the complementary fragment, or the interaction energy between them).
Thus, for the first time in the literature, we give a definition of
scission that relies on the non-local aspects of quantum mechanics.

This leads us to describe our system after scission in the Hill-Wheeler
approximation as $\Psi=\int f\left(d\right)\Phi_{d}dd$ where $d$
is the relative distance between the fragments, and $\Phi_{d}\equiv\Phi_{1}\Phi_{2}$
is the two fragments' wave function. We obtain the collective Hamiltonian
\cite{berger80},\begin{eqnarray*}
H_{\textrm{coll}} & \equiv & \frac{\vec{p}_{d}^{2}}{2\mu m}+V\left(d\right)+C\end{eqnarray*}
where $\vec{p}_{d}$ is the momentum operator corresponding to $d$,
$\mu$ is the reduced mass of the fragments with masses $A_{1}$ and
$A_{2}$, and $m$ is the nucleon mass, $V\left(d\right)$ is the
fragment interaction potential, and $C=E_{i}+\varepsilon_{0}$ is
a constant with $E_{i}$ ($i=1,2$) the internal fragment energy,
and $\varepsilon_{0}$ a zero-point correction, \begin{eqnarray*}
E_{i} & \equiv & \left\langle \Phi_{i}\left|H-\frac{\vec{p}_{i}^{2}}{2mA_{i}}\right|\Phi_{i}\right\rangle \\
\varepsilon_{0} & \equiv & \left\langle \Phi_{d}\left|\frac{\vec{p}_{1}^{2}}{2mA_{1}}+\frac{\vec{p}_{2}^{2}}{2mA_{2}}-\frac{\vec{p}^{2}}{2mA}\right|\Phi_{d}\right\rangle \end{eqnarray*}
Note that $V\left(d_{s}\right)+C$ is nothing but the total Bogoliubov
energy at the scission point (i.e., at $d=d_{s}$).

Thus we propose the following two-stage description of fission: 1)
the nucleus deforms until it reaches a scission configuration determined
with the criteria given above at which point the fragments are {}``frozen''
in their configurations and 2) as a result of their strong mutual
repulsion move apart essentially by spatial translation. Eventually
these fragments will decay by neutron and gamma emission to their
respective ground states.

In the following we investigate the extent to which this picture is
consistent with experimental observables. Let us first discuss our
predictions assuming a one-dimensional path leading to the hot fission
point and that the collective dynamic is adiabatic from the saddle
to the scission point. Starting with zero energy at the saddle we
are at $\approx$ 25 MeV above the scission point. With our assumption
this energy must be interpreted as a collective pre-kinetic energy.
Now, the kinetic energy acquired by the fragments after scission is
simply given by $V\left(d_{s}\right)-V\left(\infty\right)=V\left(d_{s}\right)$
which is $\approx$ 170 MeV according to our calculations. Adding
this pre-kinetic energy, our description gives a total kinetic energy
(TKE) of 195 MeV which exceeds by only 10 MeV the experimental value
184.8 $\pm$ 1.7 MeV obtained by averaging the data sets available
in the literature \cite{wagemans84,nishio95,tsuchiya00}.

The calculation of the fragment excitation energies requires their
corresponding ground-state energies. In order to calculate these ground-state
energies consistently within the same basis as the excited states,
they have been obtained from an HFB calculation starting from the
scission configuration, but without the constraint on neck size. Constraints
were added to keep the average number of protons and neutrons in each
fragment the same as in the excited state, but otherwise the fragments
were allowed to drop to their lowest-energy state as they were pulled
further apart. The minimum energies of the fragments were thus found
when they were moved an additional 1.6 fm apart. Using these as the
ground-state energies, excitation energies of 4.5 MeV and 7 MeV were
obtained for the heavy (average mass number $\approx$ 132) and light
(average mass number $\approx$ 108) fragment, respectively. Together,
these yield a total excitation energy (TXE) of $\approx$ 11.5 MeV.
By contrast, the average TXE expected from empirical arguments \cite{madland06}
in thermal fission on a $^{239}\textrm{Pu}$ target is $\approx$
26 MeV, thus leaving a $\approx$ 15-MeV discrepancy which we address
next. 

It is believed that three collective degrees of freedom ($Q_{20}$,
$Q_{30}$, $Q_{40}$), if not four (triaxial mode), are the barest
minimum needed to describe the collective dynamics of fission. If
so, one could expect that part of the available energy in the descent
from saddle to scission would be transferred to two or three modes
transverse to the fission direction. This possibility was studied
previously \cite{berger89} with two degrees of freedom ($Q_{20}$,
$Q_{40}$). That work showed that $\sim$ 2 MeV are already taken
by one transverse mode. With two other degrees of freedom a total
of 4 or 5 MeV could be taken up in these collective modes, at the
expense of the fragment kinetic energy. Finally an other source of
dissipation can result from the coupling of the collective dynamic
with internal excitations \cite{moyadeguerra77}. A derivation of
such coupling can be obtained in the framework of a generalization
of the generator coordinate method including two quasiparticle excitations
\cite{remi11}. Therefore we have sufficiently many degrees of freedom
to dissipate part of the available 25 MeV from saddle to scission. 

Let us therefore consider different damping scenarios. Suppose that
the 25 MeV potential energy liberated in the fission of $^{240}\textrm{Pu}$
is shared in a 50/50 split between pre-scission fragment kinetic and
excitation energies, then our prediction (TKE = 182.5 MeV and TXE
= 24 MeV) precisely matches the experimental values. Even if we take
a more conservative 25/75 distribution, one way or the other, the
scenario we propose is still in remarkable agreement with observations.
It is rather striking that, without adjustable parameters, we have
formulated a quantum mechanical and dynamical picture of scission
that is consistent with experiment.

In closing, we comment briefly on the nature of the fragment excited
states calculated in this microscopic approach. Strictly speaking,
the fragments identified on the HFB solution for $^{240}\textrm{Pu}$
should be analyzed separately on the set of eigenstates of the Hamiltonians
describing each of them. The spectrum of nuclei at such high energies
is not known and its description would likely require a statistical
distribution over all conceivable types of states at those excitation
energies (collective, intrinsic, states with one or more nucleons
in the continuum, etc.). Whether or not a statistical approach is
necessary to perform such an analysis is a separate question. The
description of fission we propose does not require any statistical
mechanics or any kind of temperature for low-energy fission.

Finally, let us mention that the concept of localization could have
interesting applications as we approach the scission point. In effect,
as we recognize pre-fragments, the values of the global constraints
split into the contributions from those pre-fragments. As the fragments
move apart, we expect the correct description of the system to rely
on separate collective coordinates for those individual fragments.
Although it remains to be verified, we believe that the localization
of Fock space could provide a way to impose constraints separately
on the pre-fragments, and thereby give a richer description of the
nucleus at and beyond scission.

This work was performed under the auspices of the US Department of
Energy by the Lawrence Livermore National Laboratory under Contract
DE-AC52-07NA27344. Funding for this work was provided by the United
States Department of Energy Office of Science, Nuclear Physics Program
pursuant to Contract DE-AC52-07NA27344 Clause B-9999, Clause H-9999
and the American Recovery and Reinvestment Act, Pub. L. 111-5.

\end{document}